# Alignment of the amplitudes of stored states for solving the associative memory problem using quantum annealing

V. E. Zobov[1)], I. S. Pichkovskiy[1)]
[1)]Kirensky Institute of Physics, Federal Research Center KSC SB RAS,
Akademgorodok 50, bld. 38, Krasnoyarsk, 660036 Russia

The solution of combinatorial optimization problems by means of quantum annealing is based on the preparation of an equally superposition state. Such superposition is crucial to the associative memory problem of correctly recalling patterns. In this paper, we study the alignment of the amplitudes of stored states in memory using the proposed auxiliary Hamiltonian, which is switched off at the end of evolution. The results of the algorithm execution were obtained both by numerical modeling and by the perturbation theory method, which made it possible to obtain the dependences on physical parameters in explicit form. As examples, systems of two and three qutrits, represented by spins $S$=1, are considered. Since the Hamiltonian is expressed in a universal form through projectors onto states stored in memory, it can be used for other quantum systems (platforms).

## 1. Introduction

There is currently a growing interest in artificial intelligence (AI). Solving AI problems cannot be done without modern physics. An example of this is Hopfield's work on associative memory [1 - 3], for which he was awarded the Nobel Prize in Physics in 2024. In the Hopfield artificial network, each binary neuron is represented as a classical spin $S=1/2$, which interacts with each other. The interactions between spins are described by the Ising model. In such a classical Hopfield network, the recording of patterns is carried out in the constants of interactions between spins, which are chosen so that each stored pattern corresponds to its own minimum energy of the system with equal energy values. The recognized pattern is fed to the input and lowers the energy of the minimum corresponding to the state to be remembered (or close to it). The recognition process is carried out by iteratively changing the states of the spins until the system ends up in input (or close to one) state.

To increase the capacity of associative memory and the speed up of pattern recall, it is proposed to move to quantum associative memory and quantum artificial neural networks (QNN) [4–11], as well as from a stationary to a dynamic version, in which memorized patterns correspond to trajectory attractors [11, 12].

Quantum annealing is considered as one of the methods for storage and recalling patterns [13 – 17]. Stored and recall patterns are represented by the target Hamiltonian. Recall is achieved through an adiabatic change of the Hamiltonian in time (quantum annealing). In this way, in the quantum case, it is possible to prepare a system not only in a state corresponding to a single pattern [13 – 15], but also in a state of their superposition of a given type [16, 17]. The quantum annealing method has proven successful in solving many optimization problems [18]. However, when preparing a superposition state, it distorts the amplitudes (weights) of the states included in the superposition [15-17, 19-21], whereas for associative memory it is necessary to prepare an equally superposition state.

Various methods have been proposed to correct the amplitudes: to modify the initial Hamiltonian [19, 21], to modify the target Hamiltonian [16, 17], to apply machine learning methods [9], and to add an auxiliary Hamiltonian that is turned off at the initial and final moments of evolution [22]. The last of the listed methods was proposed earlier [23] to increase the gap between the ground and excited states and thereby reduce the time and error of implementation. For this purpose, it is widely used today (see, for example, [24, 25]) and has been given the name “catalysis method”.

It should be noted that, in addition to quantum annealing, algorithms for preparing superposition states and extracting the required patterns using logical operators and machine learning methods have been proposed [4-11]. Nevertheless, work in this field continues. As the authors of [10] explain, each experimental situation has its own requirements, and it is unclear which method will be more successful, so it is necessary to develop various approaches.

A universal method for aligning the amplitudes of a superposition state in the quantum annealing method has not been found [21]. Its form depends on the problem being solved and the set of stored states. The auxiliary Hamiltonian proposed by us in [22] is also not universal, since it is expressed through stored states (patterns). However, the rules for obtaining it are universal, and it is symmetrical relative stored state permutation. In our previous work [22], we tested its effectiveness on systems of two and three qutrits. In the present work, we investigate the alignment mechanism. Using perturbation theory (PT) with a small parameter $\left(1-\frac{t}{T}\right)$, where $T$ is the total annealing time, formulas for the amplitudes are obtained. The PT results are compared with the results of numerical simulations for two examples: three stored patterns on a two-qutrit system and five stored patterns on a three-qutrit system. We chose qutrits because they provide greater memory capacity than qubits for the same number of neurons in the network [22], although the proposed method is also applicable to qubits.

## 2. Effective Hamiltonian and amplitudes of stored states

Let's consider a quantum artificial neural network of $n$ neurons. To implement associative memory, the stored patterns should be encoded in states of a quantum system. We take as a neuron a magnetic moment with spin $S$, and as a computational basis we will use the basis $\left|m_1, m_2, ..., m_n\right\rangle$ of eigenfunctions of the operators of projections of spins onto the Z axis. Each of the projections $m_i$ can take one of the $2S+1$ values: $S, 1, 0, -1, -S$. Consider storing $p$ patterns represented by orthogonal quantum vectors $\left|\psi_\mu\right\rangle = \left|m_1^\mu, m_2^\mu, ..., m_{\mathrm{n}}^\mu\right\rangle$, where $\mu = 1, 2, ..., p$. We choose the projection method of encoding [13, 14], in which the patterns is carried out through the memory Hamiltonian

$$H_{mem} = -\sum_{\mu=1}^{p} \left|\psi_\mu\right\rangle\left\langle\psi_\mu\right|, \tag{1}$$

where $\left|\psi_\mu\right\rangle\left\langle\psi_\mu\right|$ is the projector on state $\left|\psi_\mu\right\rangle$. The ground state of this Hamiltonian is a superposition state:

$$\left|\Psi\right\rangle = \sum_{\mu=1}^{p} a_\mu \left|\psi_\mu\right\rangle, \tag{2}$$

where $a_\mu = a(m_1^\mu, m_2^\mu, ..., m_n^\mu)$ are the state amplitudes. The recording and recall of patterns is accomplished by slowly changing the Hamiltonian over time:

$$H(t) = \frac{t}{T}\left(H_{mem} + \Gamma H_{probe}\right) + \frac{t}{T}\left(1-\frac{t}{T}\right)\varepsilon H_{help} + \left(1-\frac{t}{T}\right)H_{initial}, \tag{3}$$

where $T$ is the total evolution time, $H_{initial} = -h\sum_i S_i^x$ is the initial Hamiltonian, $H_{probe} = -\left|\psi_{prob}\right\rangle\left\langle\psi_{prob}\right|$ is the probe (stimulus) Hamiltonian of the hint, $\Gamma$ is a small constant, $\varepsilon \in [0,1]$ is a scale factor, $H_{help}$ is the auxiliary off-diagonal Hamiltonian introduced [22] to equalize the amplitudes $a_\mu$:

$$H_{help} = -\left(\sum_{\mu=1}^{p}\left|\psi_\mu\right\rangle\right)\left(\sum_{\nu=1}^{p}\left\langle\psi_\nu\right|\right) + \sum_{\mu=1}^{p}\left|\psi_\mu\right\rangle\left\langle\psi_\mu\right|. \tag{4}$$

As a result of the evolution of system (3) we obtain

$$\left|\Psi(T)\right\rangle = \hat{Q}\exp\left(-i\int_0^T H(t)\,dt\right)\left|\Psi_0\right\rangle, \tag{5}$$

where $\hat{Q}$ is the time ordering operator,

$$|\Psi_0\rangle = |S\rangle_x^{\otimes n} \tag{6}$$

is the ground state of the initial Hamiltonian, which is the direct product of the eigenvectors of the spin operators $S_x$ of individual spins with positive eigenvalues $S$. During an adiabatic change of the Hamiltonian (3), the system at each moment of time $t$ is in the instantaneous ground state of the Hamiltonian $H(t)$ (i.e., the population of this state is 1, and the rest are 0). When $t=0$ it is the state (6), and when $0<t<T$ the ground state is a time-varying combination of states of the computational basis. In the presence of the probe Hamiltonian hint in the form of a projector on the one of the stored states, the energy of this state in the effective Hamiltonian (3) decreases by $\Gamma$ at $t=T$. The energies of other states will not change. Since the degeneracy is removed and the recalled state becomes the ground state of the Hamiltonian, the system passes into this state during adiabatic passage. If $\Gamma=0$, then the system will go into the superposition state (2).

To study the mechanisms of amplitude equalization in (2), we consider the dynamics of a part of the Hilbert space of dimension $p$, the basis of which is formed by the stored states $|\psi_\mu\rangle$. At $t\to T$ the Hamiltonian (1) separates they from the upper unpopulated states by a large (of the order of 1) energy gap. In this limit, the differences in the energies of the instantaneous states of the selected subspace tend to zero, and adiabatic evolution is replaced by diabatic one [16, 17, 22, 23]. Transitions occur between the instantaneous ground state and the remaining states. We will describe the changes in the population of these states with the help of an effective Hamiltonian using PT with a small parameter $x=1-\frac{t}{T}$. The matrices of the Hamiltonians $H_{help}$, $H_{mem}$ and $H_{probe}$ in the basis of the stored states are written out simply. Since the Hamiltonian $H_{initial}=-h\sum_i S_i^x$ has non-zero matrix elements only between states with spin projections that differ by 1, higher PT orders may be needed involving transitions through upper states for the matrix of the effective Hamiltonian $\bar{H}_{initial}$. The $\bar{H}_{initial}$ derivation is described in the supplementary material [33].

Let us turn to the basis that diagonalizes $H_{help}$. Due to the symmetry under permutations of states, such a basis is obtained using the matrix of the quantum Fourier transform [26]:

$$F=\frac{1}{\sqrt{p}}\begin{bmatrix} 1 & 1 & 1 & \cdots & \\ 1 & \sigma & \sigma^2 & \cdots & \\ \vdots & \vdots & \vdots & \ddots & \vdots \\ 1 & \sigma^{p-1} & \sigma^{2(p-1)} & \ddots & ^2 \end{bmatrix} \quad = \exp\left(\frac{2\pi i}{p}\right). \tag{7}$$

After transformation we obtain

$$\begin{aligned} F^*H(t)F &= (t/T)\Gamma F^*H_{probe}F + (1-t/T)F^*H_{initial}F - \\ &-(t/T)(1-t/T)\varepsilon pD - (t/T)\left[1-\varepsilon(1-t/T)\right]E, \end{aligned} \tag{8}$$

where $E$ is the identity matrix of order $p$, and $D$ is a matrix of order $p$ with one non-zero element, the first element of the diagonal, equal to 1. This first element corresponds to the equally superposition state (2) at $a_\mu=\frac{1}{\sqrt{p}}$. At $\varepsilon=1$ and $\Gamma$=0, this term of the Hamiltonian ensures the dominance of the equally superposition state in the evolution process $t\to T$.

In the Hamiltonian (8), matrices $(1-t/T)F^*H_{initial}F$ and $(t/T)\Gamma F^*H_{probe}F$ have off-diagonal elements, which induce a transition from the ground (symmetric) state to excited states. As a result of these transitions, the amplitudes (populations) of the states will change. Let's find the time dependence of the system state in the Fourier basis:

$$\left|\Psi(t)\right\rangle = \sum_{i=1}^{i=p} a_i(t)\left|\varphi_i\right\rangle \tag{9}$$

with non-stationary PT, in which an off-diagonal perturbation matrix $V_{ji}(t)$ (8) (or $\bar{H}_{initial}$ [33] see (S.2) and (S.17)) causes transitions between levels. Energy levels $E_i(t)$ are determined by diagonal matrix elements (8). We find in the Fourier basis, which also does not change over time:

$$\left|\varphi_i\right\rangle = \sum_{\mu=1}^{p} F_{\mu i}\left|\psi_\mu\right\rangle, \ \left|\psi_\mu\right\rangle = \sum_{i=1}^{p} F_{\mu i}^{*}\left|\varphi_i\right\rangle. \tag{10}$$

To perform the estimation, we assume that $t = t_d$ is the moment in time when adiabatic evolution changes to diabatic one, and the value $a_1(t \le t_d) = 1$, but $a_i(t \le t_d) = 0$ for $i > 1$. We will discuss the validity of this approximation at the end of section 4, and now we will obtain results for the state amplitudes in this approximation. The off-diagonal perturbation elements that cause amplitude changes are of the next order of smallness in $x$ compared to the energy differences, and will therefore oscillate rapidly. Let's turn to interaction representation based on these oscillations:

$$\alpha_i(t) = \exp\left\{-i\int_{t_d}^{t} E_i(t_1)dt_1\right\}, \tag{11}$$

and represent the amplitudes in the form $a_i(t) = \alpha_i(t)\beta_i(t)$. In the linear perturbation approximation $\beta_1(t) = 1$, and for $\beta_i(t)$ ($i > 1$) we obtain:

$$\beta_i(t)\frac{d}{dt}\alpha_i(t) + \alpha_i(t)\frac{d}{dt}\beta_i(t) = -iE_i\alpha_i(t)\beta_i(t) - iV_{i1}(t)\alpha_1(t).$$

After transformation we obtain:

$$\frac{d}{dt}\beta_i(t) = -i\frac{\alpha_1(t)}{\alpha_i(t)}V_{i1}(t).$$

Integrating this equation, we find under the condition $\beta_i(t_d) = 0$:

$$\beta_i(t) = -i\int_{t_d}^{t}\frac{\alpha_1(t_1)}{\alpha_i(t_1)}V_{i1}(t_1)dt_1. \tag{12}$$

Substituting the solution (12) into (9), we obtain the time dependence for the system state. We multiply this state by the vector state $\left|\psi_\mu\right\rangle = \left|m_1^\mu, m_2^\mu, ..., m_n^\mu\right\rangle$ and we find the projection of the final state of the system onto this vector i.e. the amplitude $a_\mu = a(m_1^\mu, m_2^\mu, ..., m_n^\mu)$:

$$\left\langle\psi_\mu\middle|\Psi(t)\right\rangle = \sum_{i=1}^{i=p} a_i(t)F_{\mu i} = \alpha_1(t)\left\{F_{\mu 1} + \sum_{i=2}^{i=p} F_{\mu i}b_i(t)\right\}, \tag{13}$$

where

$$b_i(t) = \frac{\alpha_i(t)}{\alpha_1(t)}\beta_i(t) = (-i)\frac{\alpha_i(t)}{\alpha_1(t)}\int_{t_d}^{t}\frac{\alpha_1(t_1)}{\alpha_i(t_1)}V_{i1}(t_1)dt_1 = (-i)\int_{t_d}^{t}\frac{\alpha_i(t)/\alpha_i(t_1)}{\alpha_1(t)/\alpha_1(t_1)}V_{i1}(t_1)dt_1 =$$

$$= (-i)\int_{t_d}^{t}\exp\left\{-i\int_{t_1}^{t}\left[E_i(t_2) - E_1(t_2)\right]dt_2\right\}V_{i1}(t_1)dt_1.$$

At the end of the evolution at $t = T$, it is convenient to move to the variable $x = 1 - \frac{t}{T}$, $x_0 = 1 - \frac{t_d}{T}$, and rewrite (13) in the following form:

$$\left\langle \psi_\mu \middle| \Psi(T) \right\rangle = \alpha_1(T)\left\{ F_{\mu 1} + \sum_{i=2}^{i=p} F_{\mu i} b_i(T) \right\}, \tag{14a}$$

where

$$b_i(T) = (-i)T\int_0^{x_0} \exp\left\{ -iT\int_0^{x_1} \left[ E_i(x_2) - E_1(x_2) \right] dx_2 \right\} V_{i1}(x_1) dx_1 . \tag{14b}$$

The squares of the modules of the amplitudes (14a) determine the probabilities of detecting stored states. When calculating the module, the fast oscillations $\alpha_1(T)$ is vanishing and the time dependence is determined by the $b_i(T)$ (14b).

To analyze the scope of the PT results and their accuracy, we will perform numerical simulation of the implementation of associative memory. To calculate the evolution of system (5) with time-dependent Hamiltonian (3), we divide the time $T$ into $N$ small time intervals $\Delta t = 0.01$, on which the change in the Hamiltonian (3) can be neglected [14, 22]. As a result, instead of (5) we obtain:

$$\left| \Psi(\mathrm{T}) \right\rangle \cong \prod_{l=0}^{N} \exp\left\{ -i\Delta t\left( \left( 1-\frac{l}{N} \right) H_{initial} + \frac{l}{N}\left( 1-\frac{l}{N} \right) \varepsilon H_{help} + \frac{l}{N}\left( H_{mem} + \Gamma H_{probe} \right) \right) \right\} \left| \Psi_0 \right\rangle . \tag{15}$$

We consider two examples. The first is a system of two qutrits (S=1 spin) with three stored states: $\left|0,1\right\rangle$, $\left|1,0\right\rangle$, $\left|-1,-1\right\rangle$. The second is a system of three qutrits with five stored states: $\left|1,1,1\right\rangle$, $\left|1,-1,0\right\rangle$, $\left|0,0,-1\right\rangle$, $\left|-1,1,-1\right\rangle$ и $\left|-1,-1,1\right\rangle$. The results are presented in Fig. 1, Fig. 2, and Fig. 3.

## 3. System of three spins $S = 1$.

In the memory Hamiltonian $H_{mem} = -\sum_{\mu=1}^{p} \left| \psi_\mu \right\rangle \left\langle \psi_\mu \right|$ we encode five states: $\left|1,1,1\right\rangle$, $\left|1,-1,0\right\rangle$, $\left|0,0,-1\right\rangle$, $\left|-1,1,-1\right\rangle$ and $\left|-1,-1,1\right\rangle$. The initial Hamiltonian have zero matrix elements between the chosen states. Therefore, we found the matrix for the effective Hamiltonian in the second order of PT with $x = 1-\frac{t}{T}$, taking into account transitions through the upper levels [33]. Father we will retain only the leading terms in $x$. After the transformation to the Fourier basis, the perturbations $V = x\varepsilon H_{help} + x\bar{H}_{initial}$ take the form:

$$V' = -\frac{h^2 x^2}{10}\begin{bmatrix} 22 & A_1 & A_2 & A_2^* & A_1^* \\ A_1^* & 2A_3 & A_4 & A_5 & A_6 \\ A_2^* & A_4^* & 2A_7 & A_8 & A_5 \\ A_2 & A_5^* & A_8^* & 2A_7 & A_9 \\ A_1 & A_6^* & A_5^* & A_9 & 2A_3 \end{bmatrix} - \varepsilon x \begin{bmatrix} 4 & 0 & 0 & 0 & 0 \\ 0 & -1 & 0 & 0 & 0 \\ 0 & 0 & -1 & 0 & 0 \\ 0 & 0 & 0 & -1 & 0 \\ 0 & 0 & 0 & 0 & -1 \end{bmatrix}, \tag{16}$$

where $A_1 = 4(-1)^{\frac{4}{5}} + (-1)^{\frac{2}{5}} - 2(-1)^{\frac{1}{5}}$, $A_2 = (-1)^{\frac{4}{5}} - 4(-1)^{\frac{3}{5}} + 2(-1)^{\frac{2}{5}}$, $A_3 = 9 + (-1)^{\frac{2}{5}} - (-1)^{\frac{3}{5}}$, $A_4 = 3(-1)^{\frac{2}{5}} + 2(-1)^{\frac{4}{5}} - 2(-1)^{\frac{3}{5}}$, $A_5 = 3(-1)^{\frac{4}{5}} - 2(-1)^{\frac{3}{5}} - 2(-1)^{\frac{1}{5}}$, $A_6 = 2(-1)^{\frac{2}{5}} - (-1)^{\frac{1}{5}} + 4$, $A_7 = 9 + (-1)^{\frac{4}{5}} - (-1)^{\frac{1}{5}}$, $A_8 = 2(-1)^{\frac{4}{5}} + (-1)^{\frac{2}{5}} + 4$, $A_9 = 3(-1)^{\frac{2}{5}} + 2(-1)^{\frac{4}{5}} - 2(-1)^{\frac{3}{5}}$. An asterisk in a superscript denotes of a complex conjugate matrix elements.

Specifically, in case (16) in the leading order in $x$ we have in formula (14b):

$E_i(x_2) - E_1(x_2) = 5\varepsilon x_2$, $V_{i1}(x_1) = -\frac{h^2 x_1^2}{10} A_i'$, where $A_2' = A_1^*$, $A_3' = A_2^*$, $A_4' = A_2$, $A_5' = A_1$.

The formula (14b) takes the form:

$$b_i = -\frac{iTh^2 A_i'}{10}\int_0^{x_0} x^2 \exp\left[-\frac{5iT\varepsilon x^2}{2}\right]dx, \tag{17}$$

and after variable substitution $\Omega^2 = \frac{5T\varepsilon}{2}$, $y = \Omega x$, $z = y^2$ and $z_0 = (x_0\Omega)^2$:

$$b_i = -\frac{iTh^2 A_i'}{20\Omega^3}\int_0^{z_0} \sqrt{z}\exp[-iz]dz. \tag{18}$$

Using the formula for integration by parts, we divide the integral into the middle and oscillating parts: $\int_0^{z_0} \sqrt{z}\exp[-iz]dz = \frac{-i}{2}\int_0^{z_0}\frac{1}{\sqrt{z}}\exp[-iz]dz - i\sqrt{z_0}\exp[-iz_0]$. We take the limit value $z_0 \to \infty$ of the integral for estimating its average: $\int_0^{\infty}\frac{e^{-iz}}{\sqrt{z}}dz = \Gamma\left(\frac{1}{2}\right)i^{-\frac{1}{2}} = \sqrt{\pi}i^{-\frac{1}{2}}$, and discarding the oscillating part [33]. For the slowly changing parts of the amplitudes we find the result:

$$b_i = BA_i', \tag{19}$$

where $B = \frac{h^2\sqrt{5\pi}(1-i)}{500\sqrt{T}\varepsilon^{\frac{3}{2}}}$, $\beta_1 = \sqrt{1-|\beta_2|^2-|\beta_3|^2-|\beta_4|^2-|\beta_5|^2}$.

After substituting (19) into (14a) we obtain the following results:

$$\begin{gathered} a(1,1,1) = a(-1,-1,1) = \frac{\beta_1}{\sqrt{5}} - 7\frac{B}{\sqrt{5}},\ a(1,-1,0) = \frac{\beta_1}{\sqrt{5}} - 2\frac{B}{\sqrt{5}}, \\ a(0,0,-1) = \frac{\beta_1}{\sqrt{5}} + 13\frac{B}{\sqrt{5}},\ a(-1,1,-1) = \frac{\beta_1}{\sqrt{5}} + 3\frac{B}{\sqrt{5}}, \\ \beta_1 = \sqrt{1-\frac{0.00224\pi h^4}{T\varepsilon^3}}. \end{gathered} \tag{20}$$

The comparison performed in Fig. 1 showed that the results of PT (20) are in good agreement with the results of numerical simulation.

## 4. System of two spins S=1.

We place three states: $|1,0\rangle$, $|0,1\rangle$ and $|-1,-1\rangle$ into the memory Hamiltonian $H_{mem}$. The initial Hamiltonian has zero matrix elements between the chosen states. Therefore, we found the matrix for the effective Hamiltonian in the second order of PT at small $x = 1-\frac{t}{T}$ using transitions through upper levels [33].

Let's turn to the basis:

$$\begin{aligned} |\varphi_1\rangle &= \frac{1}{\sqrt{3}}|1,0\rangle + \frac{1}{\sqrt{3}}|0,1\rangle + \frac{1}{\sqrt{3}}|-1,-1\rangle, \\ |\varphi_2\rangle &= \frac{1}{\sqrt{6}}|1,0\rangle + \frac{1}{\sqrt{6}}|0,1\rangle - \frac{2}{\sqrt{6}}|-1,-1\rangle, \\ |\varphi_3\rangle &= \frac{1}{\sqrt{2}}|1,0\rangle - \frac{1}{\sqrt{2}}|0,1\rangle. \end{aligned} \tag{21}$$

The perturbation $V = x\varepsilon H_{help} + x\bar{H}_{initial}$ in this basis will take the form:

$$V = -h^2x^2\begin{bmatrix} 2 & 1/\sqrt{2} & 0 \\ 1/\sqrt{2} & 3/2 & 0 \\ 0 & 0 & 1/2 \end{bmatrix} - \varepsilon x\begin{bmatrix} 2 & 0 & 0 \\ 0 & -1 & 0 \\ 0 & 0 & -1 \end{bmatrix}. \qquad (22)$$

The state $|\varphi_3\rangle$ is independent of the other two. We took a combination (21) of Fourier basis vectors (10) to reduce the number of changing vectors to two. For the second state, the formula (14b) takes the form:

$$b_2 = \frac{iTh^2}{\sqrt{2}}\int_0^{x_0} x^2 e^{-iTx^2\varepsilon 3/2}dx\,. \qquad (23)$$

We repeat the same reasoning as in deriving (17-19), and, for the slowly changing amplitude part of the second state, we obtained:

$$b_2 = \frac{(1-i)h^2\sqrt{\pi}}{6\sqrt{6T\varepsilon^3}} \text{ and } \beta_1 = \sqrt{1-|\beta_2|^2}\,. \qquad (24)$$

For two spins, we considered the change of states when adding probe Hamiltonian. In the linear approximation in Γ, the following results were obtained for the slowly changing parts of the amplitudes of the three stored states at the end of annealing [33]:

1. Pattern $|-1,-1\rangle$ recognition with $\Gamma H_{prob} = -\Gamma|-1,-1\rangle\langle -1,-1|$:

$$a(1,0) = a(0,1) = \frac{1}{\sqrt{3}}\beta_1 + \frac{1}{\sqrt{6}}b_2,\ a(-1,-1) = \frac{1}{\sqrt{3}}\beta_1 - \frac{\sqrt{2}}{\sqrt{3}}b_2,$$
$$b_2 = \frac{(1-i)h^2}{36}\sqrt{\frac{6\pi}{T\varepsilon^3}} - \frac{(1+i)}{18}\sqrt{\frac{6\pi T}{\varepsilon}}\Gamma + \frac{\sqrt{2}h^2}{27\varepsilon^2}\Gamma,\ \beta_1 = \sqrt{1-|\beta_2|^2}. \qquad (25)$$

2. Pattern $|0,1\rangle$ recognition with $\Gamma H_{prob} = -\Gamma|0,1\rangle\langle 0,1|$:

$$a(1,0) = \frac{1}{\sqrt{3}}(\beta_1 + \frac{1}{\sqrt{2}}b_2) + \frac{1}{\sqrt{2}}b_3,\ a(0,1) = \frac{1}{\sqrt{3}}(\beta_1 + \frac{1}{\sqrt{2}}b_2) - \frac{1}{\sqrt{2}}b_3,$$
$$a(-1,-1) = \frac{1}{\sqrt{3}}\beta_1 - \frac{\sqrt{2}}{\sqrt{3}}b_2, b_2 = (1-i)\frac{\sqrt{6\pi}h^2}{36\sqrt{T\varepsilon^3}} + (1+i)\frac{\sqrt{6\pi T}}{36\sqrt{\varepsilon}}\Gamma - \frac{\sqrt{2}h^2}{54\varepsilon^2}\Gamma, \qquad (26)$$
$$b_3 = -(1+i)\frac{\sqrt{2\pi T}}{12\sqrt{\varepsilon}}\Gamma,\ \beta_1 = \sqrt{1-|\beta_2|^2-|\beta_3|^2}.$$

The comparison showed that the results of PT (24) – (26) are in good agreement with the results of numerical simulation. This is particularly evident in Fig. 2.

Based on the calculations performed, we will discuss the essence and validity of the approximation used above, involving the transition from adiabatic to diabatic evolution. Strictly speaking, the change in the populations of the levels occurs throughout the entire evolution $(0,T)$. However, the main contribution to this process comes from the immediate vicinity of the final point $T$, which we have defined as the interval $(t_d,T)$. For the description, we used the terminology from [16, 17]. We do not use the specific numerical value of $t_d$, and it does not appear in the final formulas and results (after discarding oscillations). With its help, we limit our estimates to a small neighborhood $T$, in which PT with the small parameter $x = 1-\frac{t}{T}$ is applicable. In this neighborhood, we derive a system of equations (11) – (12) for the amplitudes of states in the subspace of stored states. For an approximate analytical solution of this system, we limit ourselves to a linear approximation with respect to a perturbation that causes transitions and has a quadratic smallness in $x$ in our case. Therefore, the second-order contributions of the nonstationary PT will be of the order of $x^4$ and will not have an impact on populations. In this approxima-tion, we took the amplitude of the ground state as 1 and the amplitude of all other states as 0 as the initial conditions. For the amplitudes, we obtain results in the form of definite integrals: the final limit of integration is $T$ ($x = 0$), and the initial $t = t_d$ ($x = x_0$). Since the integrand is a rapidly growing function, the value of the integral is determined by the value of this function at the end of the integration region and practically does not depend on the starting point. (After changing the direction of integration, it does not depend on the final limit of integration). This is a well-known property of integrals with singular points (in our case, $t = T$), the magnitude of which is determined by the neighborhood of the singular point. We do not estimate the actual amplitudes of the states at $t = t_d$ (this is difficult), but instead, we perform a comparison with a numerical experiment in Figures 1–3, where good agreement is observed, indicating the smallness of the unaccounted corrections to the population of the levels outside the interval $(t_d, T)$.

## 5. Discussion

Figures 1, 2, and 3 compare the probabilities of detecting stored states. Clearly, the PT results (20), (25) and (26) are in good agreement with the numerical simulation results. In Fig. 3, in addition to the slowly varying with $T$ detection probability, which is in good agreement with PT, the numerical simulation results contain a rapidly oscillating contribution of small amplitude (T times smaller than the slowly varying part, as follows from (S.8), (S.9). The high frequency of oscillations and their small amplitude are determined by the large energy difference between the subspace of stored states and all others (taken as 1 in our case for the memory Hamiltonian (1)). As shown in the supplementary material [33], the rapidly oscillating correction is a consequence of replacing the real Hamiltonian $\left(1-\frac{t}{T}\right)H_{initial}$ in (3) with an effective one $\left(1-\frac{t}{T}\right)\bar{H}_{initial}$, obtained using the second-order stationary PT after taking into account the transition through higher levels. The small oscillations depending on $T$ are described by the second order of non-stationary PT and do not significantly affect the results.

Let us pay attention to the importance of the factor $\left(1-\frac{t}{T}\right)$ before $H_{help}$ for the implementation of associative memory. If we replace $\varepsilon\left(1-\frac{t}{T}\right)$ in (3) with 1, we obtain the memory Hamiltonian proposed in the work [13],

$$H'_{mem} = -|\Psi\rangle\langle\Psi|, \tag{27}$$

in which $|\Psi\rangle$ is defined in (2) with $a_\mu = \frac{1}{\sqrt{2}}$. At $t \to T$ the system also moves into the $|\Psi\rangle$ state. The fundamental difference in the dynamics of two systems with Hamiltonian (1) or (27) is the response to the probe Hamiltonian (see Fig. 2 in [22]). For Hamiltonian (27), the gap between $|\Psi\rangle$ and all other states remains large at $t \to T$. To recall one of the stored states from the symmetric combination, it is necessary to take $\Gamma$ of the same order as the large gap. In particular, for two spins, after finding the correction to the ground state from adding the probe Hamiltonian $\Gamma H_{prob} = -\Gamma|0,1\rangle\langle 0,1|$ to the Hamiltonian (27), we find:

$$\frac{d}{d\Gamma}|a(0,1)|^2 \simeq [illegible] \frac{[illegible]}{\sqrt{3}}) \approx 0,3. \tag{28}$$

In the case of the Hamiltonian (8), the gap between the different stored states $\varepsilon(p-1)(1-t/T)$ disappears at $t \to T$. Therefore, a small value of $\Gamma$ is sufficient to eliminate the degeneracy. For example, for $\varepsilon = 1$, $h = 1$, $T = 350$ in (27) we obtain:

$$\frac{d}{d\Gamma}|a(0,1)|^2 \simeq [illegible] \sqrt{\frac{T}{\varepsilon}} \approx 4{,}3 . \qquad (29)$$

At the same time, the gap between the stored and unstored states remains large. Thus, the proposed model (bypasses) solves the problem of “excessive shift” identified by Seddiki et al. (reference [14]). Small $\Gamma$ is sufficient to recall the required state from the recorded (memorized) ones, but insufficient to bring the system to the input state itself if it is not among the recorded ones.

Thus, we obtain an estimate of the probabilities of detecting stored states in the form of formulas containing an explicit dependence on the parameters $\varepsilon$, $h$, and $T$, using perturbation theory. There is qualitative and quantitative agreement between these results and the results of numerical modeling of the solution to the associative memory problem. Based on the comparison and analysis of these results, we can draw the following conclusion about the mechanism for equalizing the amplitudes of the stored states using the auxiliary Hamiltonian $H_{help}$ (4) proposed by us.

At $t = 0$, the ground state $H_{initial}$ in the transverse field forms a superposition of all states in the computational basis. During the evolution without $H_{help}$, each of these states transitions into states with different mixing rates (due to different Hamming distances), which leads to an uneven superposition at $t = T$. If we include $H_{help}$, the superposition of all states in the computational basis transitions into new (instantaneous) states of effective Hamiltonian, among which the ground state is now an equally superposed state in the subspace of memorized states. When we transit to the so-called diabatic regime, $H_{help}$ provides a gap $p\varepsilon x$ between this state and the other states of the effective Hamiltonian subspace. The computational basis states that are not included in the subspace of stored states are still separated by a large energy difference $\frac{t}{T}$. The gap $p\varepsilon x$ approaches zero as $t \to T$, following $x = 1 - \frac{t}{T}$, while the rates of symmetry breaking transitions $H_{initial}$ behave as $x^2$. This explains the $H_{help}$ and $H_{initial}$ Hamiltonians competition: the amplitude difference decreases with increasing $\varepsilon$ as $\varepsilon^{-\frac{3}{2}}$ and with decreasing $h$ as $h^2$. However, according to (29), as ε increases, the sensitivity to the hint decreases (as $\varepsilon^{-1/2}$). In other words, the Hamiltonian $H_{help}$, which is no diagonal in the computational basis and symmetric with respect to transitions between stored states, ensures fast equalizing transitions between the recorded states. Thus, the Hamiltonian $H_{help}$ corrects the distortions introduced by the Hamiltonian $H_{initial}$, which is asymmetric with respect to such transitions, effectively performing an iterative update of the qubit interaction weights in the aforementioned QNN model [8, 9]. In this way, the proposed auxiliary Hamiltonian eliminates the need for iterative training processes.

Therefore, using the auxiliary Hamiltonian $H_{help}$ (4), it is possible to align the amplitudes of stored states when implementing associative memory through quantum annealing. Like the memory Hamiltonian $H_{mem}$ (1), this Hamiltonian is introduced using projectors onto the states selected for recording. Diagonal projectors form $H_{mem}$, and off-diagonal ones form $H_{help}$. The technical solution to the problem of their creation depends on the quantum system (platform) used to implement quantum algorithms on qutrits: NV-centers in diamond (paramagnetic color centers) [27], neutral (Rydberg) atoms [28], ions in traps [29], superconductors [30], etc. Projector implementation methods have been developed for each platform. For example, for a system of spins *S*=1, we found sequences of electromagnetic field pulses, resonant with transitions between levels, separated by evolution intervals, to solve problems [31]. In particular, a method for obtaining three-spin interactions has been proposed [32]. Similar work is being carried out for other platforms.

**Acknowledgments**

The study was supported by a grant from the Russian Science Foundation No. 25-22-00750, https://rscf.ru/en/project/25-22-00750/.

## Figures:

Fig. 1. Dependences of the probabilities of detecting stored states for 3 spins on ε for $h=1$ (a) and on $h$ for $\varepsilon=1$ (b) for $\Gamma=0$ and $T=350$. Solid lines from top to bottom show the results of numerical simulation for states $|0,0,-1\rangle$, $|-1,1,-1\rangle$, $|1,-1,0\rangle$, $|-1,-1,1\rangle$, and $|1,1,1\rangle$ (the last two coincided). Dashed lines show the results of PT (20) for the corresponding states.

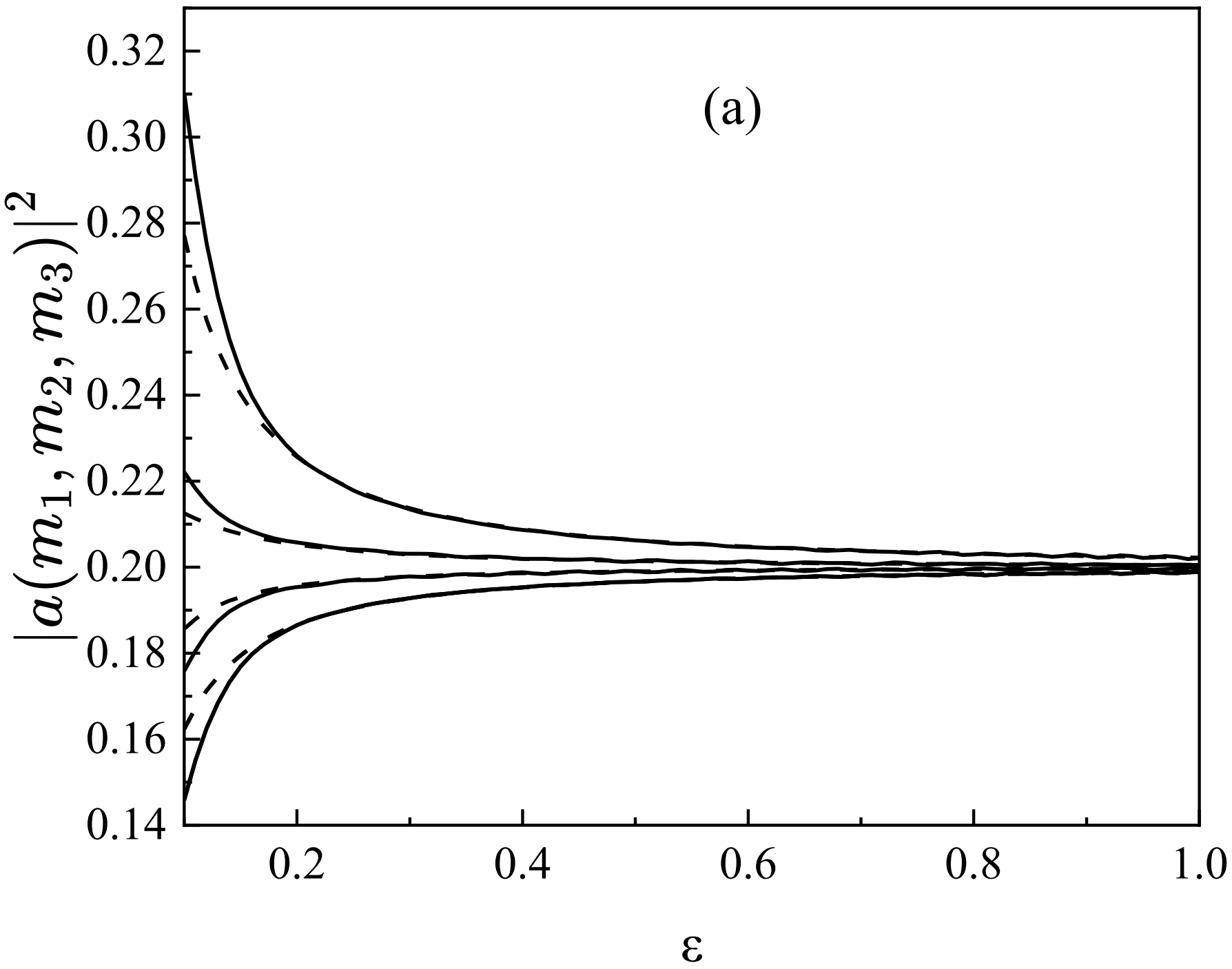

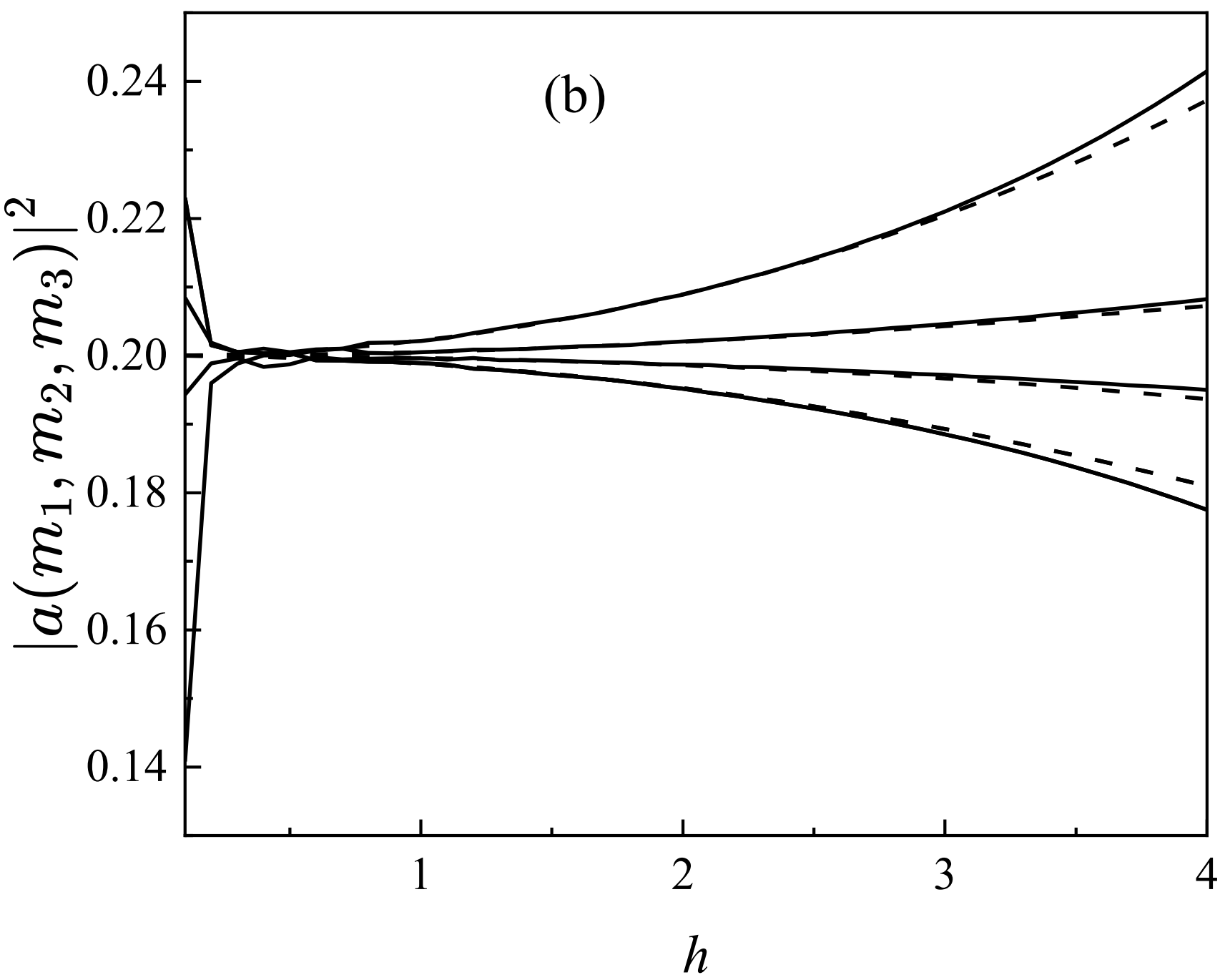

Fig. 2. Dependences on ε of the derivatives with respect to Γ (at $\Gamma=0$) of the probabilities of detecting stored states for two spins: (a) with $\Gamma H_{prob}=-\Gamma|-1,-1\rangle\langle-1,-1|$ and (b) with $\Gamma H_{prob}=-\Gamma|0,1\rangle\langle0,1|$. The values $\Gamma H_{prob}=-\Gamma|-1,-1\rangle\langle-1,-1|$ of other parameters are: $h=1$ and $T=350$. The solid lines show the results of numerical simulation. The dashed lines show the results of PT for the corresponding states.

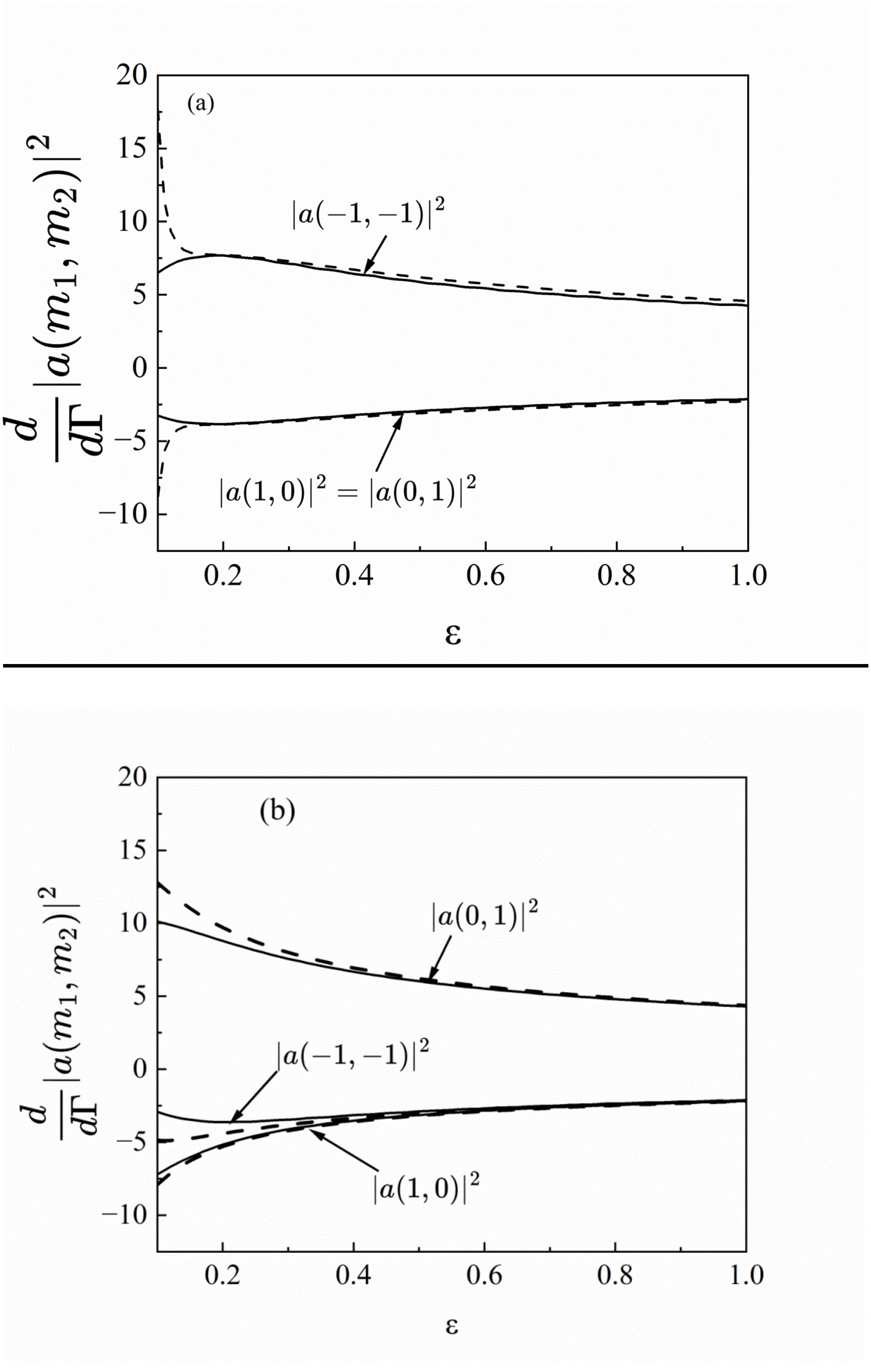

Fig. 3. Dependences of the probabilities of detecting stored states on $T$ for 2 spins (a) and for 3 spins (b), for $h=1$, $\varepsilon=1$ and $\Gamma=0$. Solid lines show the results of numerical simulation of the probabilities of detecting stored states: a) from top to bottom for states $|1,0\rangle$, $|0,1\rangle$ and $|-1,-1\rangle$ (the first two coincided), b) from top to bottom for states $|0,0,-1\rangle$, $|-1,1,-1\rangle$, $|1,-1,0\rangle$, $|-1,-1,1\rangle$ and $|1,1,1\rangle$ (the last two coincided). Dashed lines show the PT results for the corresponding states.

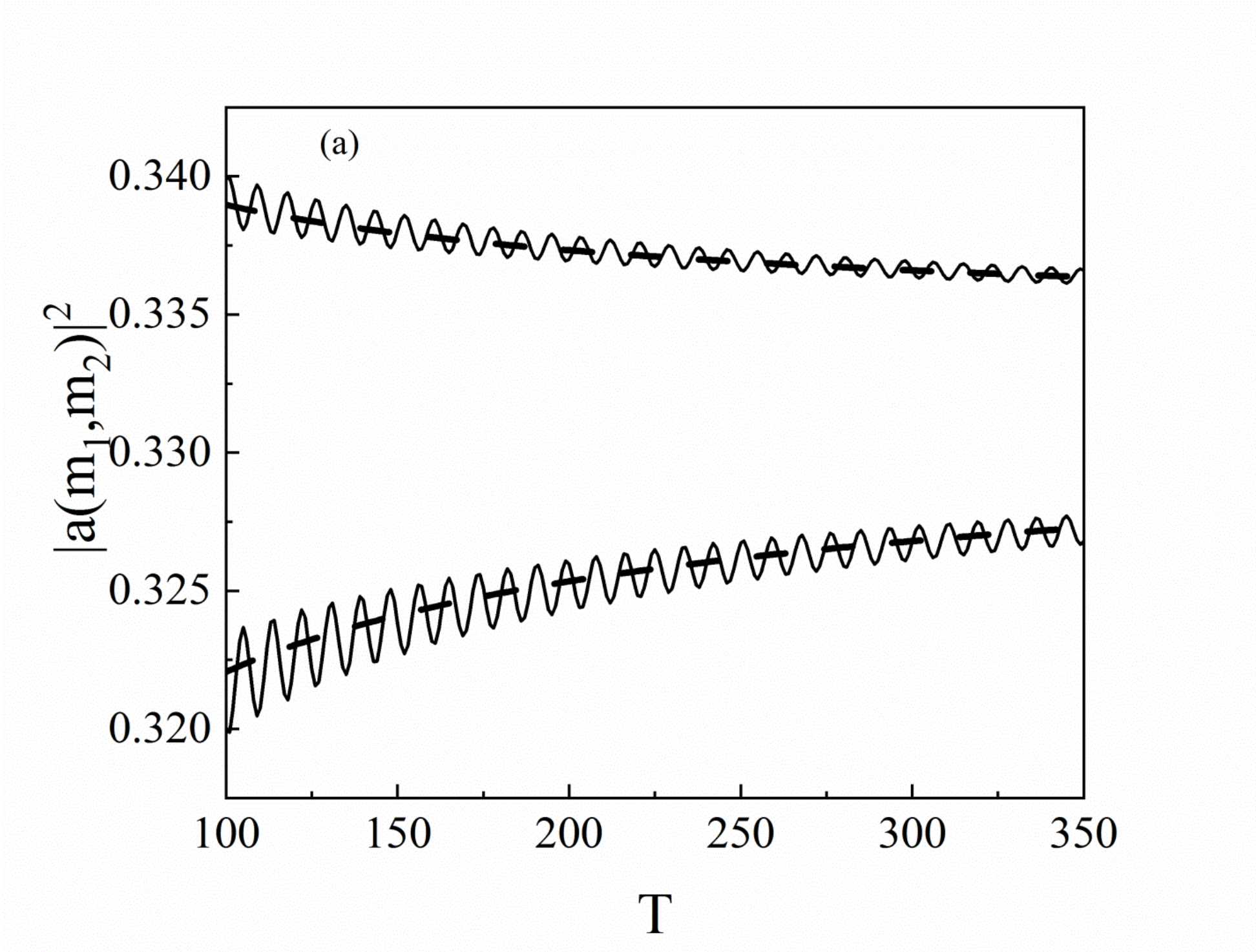

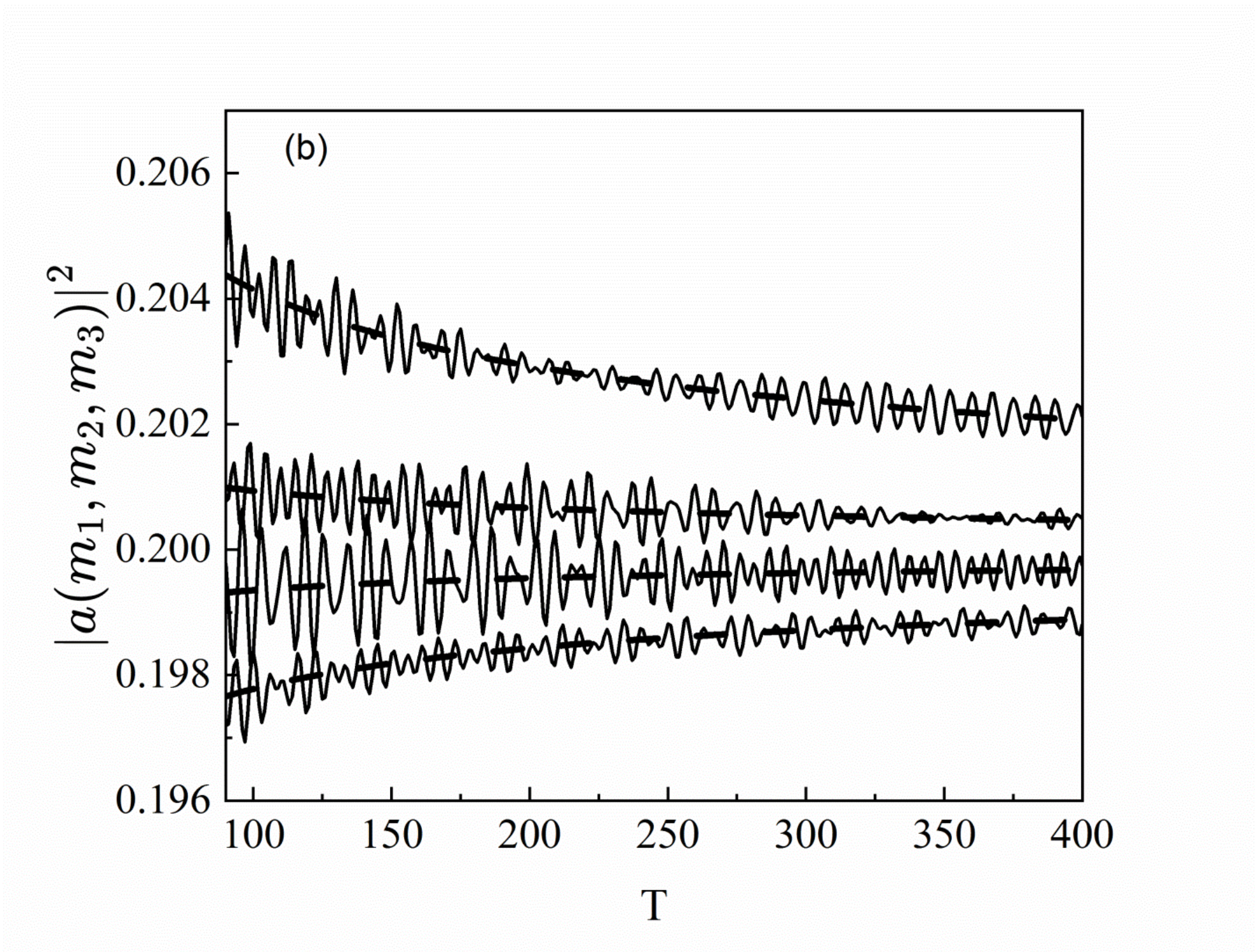

**Supplementary material** to article "Alignment of the amplitudes of stored states for solving the associative memory problem using quantum annealing",

V. E. Zobov, I. S. Pichkovskiy.

In this supplementary material, we describe the derivation of the effective Hamiltonian and the formulas given in the main text of the paper for the probabilities of detecting stored states when implementing associative memory using quantum annealing on two and three qutrits. We use the stationary and non-stationary perturbation theories (PT).

## System of two spins $S = 1$

1. **Derivation of the effective Hamiltonian.** In the memory Hamiltonian $H_{mem} = -\sum_{\mu=1}^{p} |\psi_\mu\rangle\langle\psi_\mu|$ we placed three states: $|1,0\rangle$, $|0,1\rangle$ and $|-1,-1\rangle$. The Hamiltonian $H_{initial} = -h\left(S_1^x + S_2^x\right)$, has zero matrix elements between the chosen stored states. Therefore, we find the matrix of the effective Hamiltonian in the second order of PT at a small quantity $x = 1 - \frac{t}{T}$ using transitions through the upper levels. We apply the formula for the stationary PT:

$$V'_{ij} = V_{ij} + \sum_m \frac{V_{im}V_{mj}}{E_i^{(0)} - E_m^{(0)}} \tag{S.1}$$

to the instantaneous Hamiltonian $V = xH_{initial}$ and find the effective Hamiltonian

$$x\bar{H}_{initial} = \frac{h^2x^2}{-1+x}\begin{bmatrix} 3/2 & 1 & 0 \\ 1 & 3/2 & 0 \\ 0 & 0 & 1 \end{bmatrix}. \tag{S.2}$$

In the future, when performing estimates, we will retain only the leading terms in $x$.

The matrix (S.2) has been obtained in the basis of the stored states. In the main text, we transformed this effective Hamiltonian (22) into the basis of collective states (21) and estimated the amplitude (23) of the transition $|\varphi_1\rangle \to |\varphi_2\rangle$ between level $|\varphi_1\rangle = \frac{1}{\sqrt{3}}|1,0\rangle + \frac{1}{\sqrt{3}}|0,1\rangle + \frac{1}{\sqrt{3}}|-1,-1\rangle$ with energy $E_1(x) = -(1-x)(1+2\varepsilon x)$ and level $|\varphi_2\rangle = \frac{1}{\sqrt{6}}|1,0\rangle + \frac{1}{\sqrt{6}}|0,1\rangle - \frac{2}{\sqrt{6}}|-1,-1\rangle$ with energy $E_2(x) = -(1-x)(1-\varepsilon x)$. On the other hand, we can apply non-stationary PT to estimate this transition amplitude as two successive transitions $|\varphi_1\rangle \to |\psi_k\rangle \to |\varphi_2\rangle$ under perturbation $V(t) = -hx\left(S_1^x + S_2^x\right)$ through the upper levels: $|0,0\rangle, |1,1\rangle, |-1,1\rangle, |1,-1\rangle, |-1,0\rangle, |0,-1\rangle$ with energy $E_k = 0$.

In the second order of PT instead of (12) we get

$$\beta_2(t) = -\sum_k \int_{t_d}^{t} dt_2 \frac{\alpha_k(t_2)}{\alpha_2(t_2)} V_{2k}(t_2) \int_{t_d}^{t_2} dt_1 \frac{\alpha_1(t_1)}{\alpha_k(t_1)} V_{k1}(t_1). \tag{S.3}$$

Since $E_k = 0$, then $\alpha_k(t) = 1$ and the summation over transitions is performed independently of the integration over time. For example, the amplitudes of transitions through levels $|0,0\rangle$ and $|0,-1\rangle$ are

$$\langle\varphi_1|V(t_1)|0,0\rangle\langle 0,0|V(t_2)|\varphi_2\rangle = x_1x_2h^2\frac{\sqrt{2}}{3},\ \langle\varphi_1|V(t_1)|0,-1\rangle\langle 0,-1|V(t_2)|\varphi_2\rangle = x_1x_2h^2\frac{-1}{3\sqrt{2}}.$$

Summing up over all six transitions, we find

$$\sum_k\langle\varphi_1|V(t_1)|\psi_k\rangle\langle\psi_k|V(t_2)|\varphi_2\rangle = x_1x_2h^2\frac{1}{\sqrt{2}}. \tag{S.4}$$

Substituting this result into (S.3) we obtain $\beta_2(t)$, and then the value introduced in (13) – (14):

$$b_2(T) = \frac{\alpha_2(T)}{\alpha_1(T)}\beta_2(T) = -\int_{t_d}^{T}dt_1\int_{t_1}^{T}dt_2\frac{\alpha_1(t_1)}{\alpha_1(t_2)}\frac{\alpha_2(T,t_2)}{\alpha_1(T,t_2)}x_1x_2h^2\frac{1}{\sqrt{2}}, \tag{S.5}$$

where

$$\frac{\alpha_1(t_1)}{\alpha_1(t_2)} = \exp\left\{-i\int_{t_1}^{t_2}(1-x_3)(1+2\varepsilon x_3)dt_3\right\} \tag{S.6}$$

$$\frac{\alpha_2(T,t_2)}{\alpha_1(T,t_2)} = \frac{\alpha_2(T)/\alpha_2(t_2)}{\alpha_1(T)/\alpha_1(t_2)} = \exp\left\{-i\int_{t_2}^{T}\left[E_2(t_3)-E_1(t_3)\right]dt_3\right\} = \\ = \exp\left\{-i\int_{t_2}^{T}(1-x_3)3\varepsilon x_3dt_3\right\}. \tag{S.7}$$

In (S.5) the integrand is the product of a rapidly oscillating function $\frac{\alpha_1(t_1)}{\alpha_1(t_2)} \approx \exp\{-i(t_2-t_1)\}$ and a slowly varying function (we retain only the leading terms in *x*). Let's apply the integration by parts formula $\int udv = uv - \int vdu$ with integration over the variable $t_2$ and with

$$v = i\exp\{-i(t_2-t_1)\},\qquad du = \frac{d}{dt_2}\left\{\frac{\alpha_2(T,t_2)}{\alpha_1(T,t_2)}x_2\right\}dt_2.$$

The integral (S.5) is divided into a sum of two parts. The first part

$$h^2\frac{i}{\sqrt{2}}\int_{t_d}^{T}dt_1\frac{\alpha_2(T,t_1)}{\alpha_1(T,t_1)}(x_1)^2 \tag{S.8}$$

is represents the result (23) for the transition amplitude obtained using the effective Hamiltonian (22). The second part gives a correction:

$$\frac{-ih^2}{T\sqrt{2}}\int_{t_d}^{T}dt_1\int_{t_1}^{T}dt_2\exp\{-i(t_2-t_1)\}x_1 = \frac{h^2}{\sqrt{2}}\left\{\left[\frac{ix_0}{T}+\frac{1}{T^2}\right]\exp\{-i(T-t_d)\} - \frac{x_0^2}{2}+\frac{1}{T^2}\right\}. \tag{S.9}$$

This correction oscillates with time $T$: the oscillation period is $2\pi/x_0$, if $x_0$ does not change with *T*, or $2\pi$, if $t_d$ does not change with $T$.

**2. Pattern $|-1,-1\rangle$ recognition.** In this case, the perturbation takes the form.

$$V = x\varepsilon H_{help} + x\bar{H}_{initial} - \frac{t}{T}\Gamma|-1,-1\rangle\langle -1,-1| \tag{S.10}$$

where $x\varepsilon H_{help} + x\bar{H}_{initial}$ is given in (22), and

$$-\frac{t}{T}\Gamma|-1,-1\rangle\langle-1,-1| = -(1-x)\frac{\Gamma}{3}\begin{bmatrix} 1 & -\sqrt{2} \\ -\sqrt{2} & 2 \end{bmatrix}. \tag{S.11}$$

We find the slowly changing part of the amplitude using formula (14b) after substituting (S.10) into it. The diagonal part of (S.11) will enter as an addition to the energy $E_i$, and the off-diagonal part of (S.11) will be added to $V_{i1}$. In the linear approximation with respect to Γ, repeating the same reasoning as in the derivation of (17) – (19) and (23), for the slowly changing part of the amplitude we find the expression:

$$b_2 = -\frac{iT\Gamma\sqrt{2}}{6\omega}\int_0^{z_0}\frac{e^{-iz}}{\sqrt{z}}dz + \frac{iT\sqrt{2}h^2}{4\omega^3}\int_0^{z_0}\sqrt{z}e^{-iz}dz - \frac{T^2\sqrt{2}h^2\Gamma}{12\omega^4}\int_0^{z_0} ze^{-Iz}dz\,, \tag{S.12}$$

where $\omega^2 = \frac{3}{2}T\varepsilon$. To estimate the average values of the integrals in (S.12) we take their limit values at $z_0 = (x_0\omega)^2 = \infty$:

$$\int_0^{\infty}\frac{e^{-iz}}{\sqrt{z}}dz = \sqrt{\pi}i^{-\frac{1}{2}},\quad \int_0^{\infty}\sqrt{z}e^{-iz}dz \sim \qquad \int_0^{\infty} ze^{-iz}dz \sim \qquad , \tag{S.13}$$

obtained after discarding the oscillatory contributions. The result for $b_2$ is given in formula (25) of the main text. Then, returning to the computational basis according to formula (14a), we obtain the amplitudes of the three stored states, also given in this formula (25).

When evaluating the integrals that characterize the slow changes in the populations of the levels, in (S.13) and (18) we discard the oscillating contributions with oscillation frequencies $z_0 = pT\varepsilon x_0^2/2$. We have verified that after discarding the contributions oscillating at the frequency $Z_0$, the changes in the populations are consistent with the numerical experiment shown in Figs. 1–3. Although this is not a matter of mathematical rigor: it is rather an empirical fact that can be interpreted as follows. One $t_d$ point is a rough idealization. In reality, transitions can occur within the interval ($t_{d_1}$, $t_{d_2}$), which should be averaged. After averaging, the oscillations will disappear (weaken).

3. **Pattern $|0,1\rangle$ recognition.** In this case, the perturbation takes the form

$$V = x\varepsilon H_{help} + x\bar{H}_{initial} - \frac{t}{T}\Gamma|0,1\rangle\langle 0,1|,$$

where
$$-\frac{t}{T}\Gamma|0,1\rangle\langle 0,1| = \frac{\Gamma}{6}(x-1)\begin{bmatrix} 2 & \sqrt{2} & -\sqrt{6} \\ \sqrt{2} & 1 & -\sqrt{3} \\ -\sqrt{6} & -\sqrt{3} & 3 \end{bmatrix}. \tag{S.14}$$

Unlike the previous case, a transition $|\varphi_1\rangle \leftrightarrow |\varphi_3\rangle$ has been added, which will change the amplitude of state 3. For its slowly changing part (14b) we find in the manner described above:

$$b_3 = -\frac{\sqrt{6}iT\Gamma}{12\omega}\int_0^{z_0}\frac{e^{-iz}}{\sqrt{z}}dz. \tag{S.15}$$

Taking into account $z_0 \to \infty$ (S.13), we obtain the result given in formula (26) of the main text.

For the transition $|\varphi\rangle \leftrightarrow |\varphi_2\rangle$ we obtain an estimate for the slowly changing part (14b) of the amplitude of state 2:

$$b_2 = \frac{\sqrt{2}iT\Gamma}{12\omega}\int_0^{z_0}\frac{e^{-iz}}{\sqrt{z}}dz + \frac{\sqrt{2}h^2 iT}{4\omega}\int_0^{z_0}\left(\frac{\sqrt{z}}{\omega^2} - \frac{izT\Gamma}{6\omega^3}\right)e^{-iz}dz. \quad (S.16)$$

In the limit $z_0 \to \infty$ (S.13) we obtain the result presented in the formula (26) of the main text.

## System of three spins $S = 1$

In the memory Hamilton $H_{mem}$ we encoded five states: $|1,1,1\rangle$, $|1,-1,0\rangle$, $|0,0,-1\rangle$, $|-1,1,-1\rangle$ and $|-1,-1,1\rangle$. The Hamiltonian $\left(1-\frac{t}{T}\right)H_{initial}$ has zero matrix elements between the chosen states. Therefore, we find the matrix of the effective Hamiltonian in the second order of the PT at a small quantity $x = 1 - \frac{t}{T}$ with the involvement of transitions through the upper levels according to the formula (S.1):

$$x\bar{H}_{initial} = -\frac{h^2x^2}{2}\begin{bmatrix} 3 & 0 & 0 & 0 & 0 \\ 0 & 4 & 0 & 0 & 0 \\ 0 & 0 & 5 & 2 & 0 \\ 0 & 0 & 2 & 3 & 0 \\ 0 & 0 & 0 & 0 & 3 \end{bmatrix}. \quad (S.17)$$

After the transformation in the Fourier basis the matrix of the perturbation $V = \frac{t}{T}\left(1-\frac{t}{T}\right)H_{help} + \left(1-\frac{t}{T}\right)\bar{H}_{initial}$ will take the form given in formula (16) of the main text.

The resulting effective Hamiltonian made it possible to calculate the slowly changing parts of the amplitudes of transitions between stored states. On the other hand, we can apply the non-stationary PT and consider two successive transitions $|\varphi_1\rangle \to |\psi_k\rangle \to |\varphi_i\rangle$ through upper levels. In this way, we obtain oscillatory corrections to the result of the average effective Hamiltonian, which are observed in Fig. 3 in the dependences obtained by numerical simulation. For three spins, the calculations are performed in the same way as above for two spins. In formula (S.5) it is necessary to replace $V_{21} = x_1x_2h^2\frac{1}{\sqrt{2}}$ with $V_{i1}(x_1,x_2) = \frac{h^2x_1x_2}{10}A_i'$ and take the functions under the integral in the corresponding form:

$$\begin{gathered} \frac{\alpha_1(t_1)}{\alpha_1(t_2)} = \exp\left\{-i\int_{t_1}^{t_2}(1-x_3)(1+4\varepsilon x_3)dt_3\right\}_i, \\ \frac{\alpha_i(T,t_2)}{\alpha_1(T,t_2)} = \frac{\alpha_i(T)/\alpha_i(t_2)}{\alpha_1(T)/\alpha_1(t_2)} = \exp\left\{-i\int_{t_2}^{T}\left[E_i(t_3)-E_1(t_3)\right]dt_3\right\} = \exp\left\{-i\int_{t_2}^{T}(1-x_3)5\varepsilon x_3 dt_3\right\}. \end{gathered} \quad (S.18)$$

## System of $n$ spins $S = 1$.

The perturbation causing the changes is $H_{initial}$, and to describe them, we construct effective Hamiltonian. In our case, the contribution of $H_{initial}$ is determined in the second order of the stationary PT (S.1). For the matrix element between the memorized states $|\psi_{p_1}\rangle$ and $|\psi_{p_2}\rangle$, we obtain

$$H_{p1,p2} = \sum_{m}^{2^n - p} \frac{V_{p1m} V_{mp2}}{E_{p1}^{(0)} - E_m^{(0)}} \cong -\frac{T}{t} \left\langle \psi_{p1} \left| V^2 \right| \psi_{p2} \right\rangle . \qquad \text{(S.19)}$$

When deriving this formula, we took into account that the denominator $E_{p1}^{(0)} - E_m^{(0)} = -t / T$ can be taken out from under the summation sign, and for the memorized states $V_{p_1 p_2} = 0$. Therefore, the summation can be extended to a summation over all $2^n$ states. The operator $V^2 = h^2 S_x^2 = h^2 \{S^+ S^+ + S^+ S^- + S^- S^+ + S^- S^-\} / 4$ has 1 or 2 non zero contributions to the matrix element $H_{p_1, p_2}$, regardless of the number of spins *n*. Consequently, as *n* increases, the second order perturbation approximation will not be destroyed.